\documentclass[superscriptaddress,nofootinbib,notitlepage]{revtex4-1}
\usepackage{amsfonts}
\usepackage[bookmarksopen,colorlinks]{hyperref}
\usepackage{graphicx}
\usepackage{dcolumn}
\usepackage{bm}
\usepackage{amssymb,amsmath}
\usepackage[dvips]{color}

\begin{document}

\title{{\Large McVittie solution in $f(T)$ gravity }}
\author{Cecilia Bejarano}
\email{cbejarano@iafe.uba.ar}
\author{Rafael Ferraro}
\email{ferraro@iafe.uba.ar}
\author{Mar\'ia Jos\'e Guzm\'an}
\email{mjguzman@iafe.uba.ar}

\begin{abstract}
We show that McVittie geometry, which describes a black hole
embedded in a FLRW universe, not only solves Einstein equations but
also remains as a non-deformable solution of $f(T)$ gravity. This
search for GR solutions that survive in $f(T)$ gravity is
facilitated by a null tetrad approach. We also show that flat FLRW
geometry is a consistent solution of $f(T)$ dynamical equations not
only for $T=-6H^{2}$ but also for $T=0$, which could be a
manifestation of the additional degrees of freedom involved in
$f(T)$ theories.
\end{abstract}

\maketitle

\affiliation{Instituto de Astronom\'ia y F\'isica del
Espacio (IAFE, CONICET-UBA), Casilla de Correo 67, Sucursal 28, 1428
Buenos Aires, Argentina.}

\affiliation{Instituto de Astronom\'ia y F\'isica del
Espacio (IAFE, CONICET-UBA), Casilla de Correo 67, Sucursal 28, 1428
Buenos Aires, Argentina.}
\affiliation{Departamento de F\'isica,
Facultad de Ciencias Exactas y Naturales, Universidad de Buenos
Aires, Argentina.}

\affiliation{Instituto de
Astronom\'ia y F\'isica del Espacio (IAFE, CONICET-UBA), Casilla de
Correo 67, Sucursal 28, 1428 Buenos Aires, Argentina.}



\section{Introduction}

\label{sec:intro}

It is well known that Riemann-Cartan spacetime, the underlying arena
of Einstein-Cartan gravity, is a geometry possessing non-vanishing
torsion and curvature \cite{Car22, Heh95, Bla13,Ortin}. Einstein's
general relativity (GR) is defined in a spacetime with zero torsion
and non-vanishing curvature. But there is another way of simplifying
the Riemann-Cartan spacetime: by imposing the vanishing of the
curvature tensor and letting the torsion undetermined. In this case,
the spacetime geometry is completely described in terms of the
vierbein (tetrad) field, since both the metric and the connection
depend on it. The connection is called Weitzenb\"{o}ck connection and
has torsion but vanishing curvature. For a certain choice of a
Lagrangian quadratic in this torsion, the theory reduces to the
so-called Teleparallel Equivalent of General Relativity (TEGR)
\cite{Ein}, whose dynamical equations are equivalent to those of GR
\cite{Hay79, Mal94, Arc04, Mal13, Per14, Fer16}. A decade ago, a
novel approach to modified gravity was proposed by using TEGR as a
starting point: the so-called $f(T)$ gravity \cite{Fer07, Fer12},
which mimics the proposal of $f(R)$ gravity by extending the
Lagrangian through an arbitrary function. From the very beginning,
the modified teleparallel gravity approach attracted attention since
it successfully describes an inflationary scenario without the
introduction of any inflaton field \cite{Fer07, Fer08}, together
with an explanation for the accelerated expansion of our universe
without dark energy \cite{Ben09}. Moreover, the action of $f(T)$
gravity contains only first derivatives of the dynamical variables
so that the dynamical equations are of second order, which is an
unusual feature in modified gravity. Soon enough, it was also
established that the action is not local Lorentz
invariant \cite{Fer07,Li11a,Li11b,Sot11} which relates to the fact that $%
f(T) $ gravity presents additional degrees of freedom, although their
physical nature is not yet well understood \cite{Li11a, Sot11,Li11b,Li11,
Yan11}. A lot of work has been done from then on (e.~g.~\cite{Lin10, Bam10,
Wu10, Bam11, Ben11, Fer11a, Fer11b, Li11a, Li11b, Sot11, Li11, Yan11, Wei11,
Wu11a, Wu11b, Zhe11, Tam12, Wu12, Izu13, Fio14, Bej15, Fer15, Wri16, Bah16,
Bah17}, among others). Following the steps of $f(T)$ gravity as an extension
of TEGR, it is worth mentioning that other alternative theories in
teleparallel framework were also developed in Born-Infeld \cite{Fer08,
Fer10, Fio13, Fio16a, Fio16b}, Kaluza-Klein \cite{Gen14}, or Lovelock \cite%
{Gon15} schemes.

In this work, we study the McVittie solution \cite{McVittie}, which
describes a black hole embedded in a FLRW cosmology, in the context of
modified teleparallelism. The McVittie spacetime is regular everywhere on
and outside the black hole horizon, and also away from the big bang
singularity, when the cosmology is dominated at late times by a positive
cosmological constant. The spherically symmetric geometry is parameterized
by a function $a(t)$ and a constant mass parameter $m$. Of course, it
reduces to FLRW cosmology with factor scale $a(t)$ at large radius, and to a
black hole with mass $m$ for proper limits. An exhaustive study of McVittie
spacetime was developed in \cite{Nolan} (see also Ref.~\cite{Sakai}). It was
pointed out some misleading conceptions about the geometrical interpretation
of the solutions that are shown in \cite{Kaloper}, where the McVittie
geometry and its casual structure are thoroughly analyzed. It is fair to say
that McVittie solution does not describe astrophysical black holes because
missing ingredients such as accretion and rotation are not considered.
However it is still interesting as a first approach to study compact objects
in an expanding universe.

Due to the lack of local Lorentz invariance, looking for dynamical solutions
in $f(T)$ gravity could be an awkward task. The symmetries of the metric
does not throw light on the path one must follow. In \cite{Bej15} we have
shown that a null tetrad approach seems to be a useful way to find solutions
with spherical or axial symmetry. We will apply the same strategy here.

The outline of the paper is the following. In Section \ref{sec:tele} we
briefly introduce teleparallel gravity and the so called $f(T)$ gravity. In
Section \ref{sec:mcv-geo} we present the McVittie spacetime. In Section \ref%
{sec:mcv-sol} we look for McVittie-like solutions in $f(T)$ gravity by
taking advantage of the null tetrad approach, which provides a simple way to
find suitable frames. In Section \ref{sec:cosmo} we show that FLRW cosmology
consistently solves the $f(T)$ equations for two different tetrads not
linked by a symmetry transformation of the equations. Finally, we state the
conclusions in Section \ref{sec:disc}.


\section{Teleparallel framework}

\label{sec:tele}

Teleparallelism is a framework to describe gravity where the role of the
metric tensor $g_{\mu \nu }$ is taken by the tetrad or \textit{vierbein} $\{%
\mathbf{e}_{a}(x^{\mu })\}$, $a=0,1,2,3$. The tetrad is a set of four
vectors at each point of the spacetime which are linked to the metric by the
condition of orthonormality $\mathbf{e}_{a}\cdot \mathbf{e}_{b}\ =\ \eta
_{ab}$, where $\eta _{ab}=$diag$(1,-1,-1,-1)$ is the Minkowski symbol. In
coordinate bases, the tetrad and its dual co-tetrad $\{\mathbf{e}^{a}(x^{\mu
})\}$ write as $\mathbf{e}_{a}\ =\ e_{a}^{\ \mu }\,\partial _{\mu }$ and $%
\mathbf{e}^{a}\ =\ e_{\ \mu }^{a}\,dx^{\mu }$, where $e_{a}^{\ \mu }$ and $%
e_{\ \mu }^{a}$ are the components of the tetrad and its inverse which
accomplish duality relationships ($e_{\ \mu }^{a}\,e_{b}^{\ \mu }\ =\ \delta
_{b}^{a}$ and $e_{\ \mu }^{a}\,e_{a}^{\ \nu }\ =\ \delta _{\mu }^{\nu }$).
By combining duality and orthonormality, one can get the metric tensor of
the manifold in terms of a tetrad in the tangent space,
\begin{equation}
g_{\mu \nu }\ =\eta _{ab}\ \,e_{\ \mu }^{a}\,e_{\ \nu }^{b}\ .\
\label{eq:orthonormal}
\end{equation}%
Of course, this relationship is invariant under (local) Lorentz
transformations of the tetrad. As seen, we have used Greek indices to label
coordinates and tensor components in coordinate basis; instead Latin indices
$a,b,...=0,1,2,3$ label the vectors taking part in the tetrad, and tensor
components with respect to these Lorentzian frames. These indices are
lowering and raising by metric tensor and the Minkowski symbol respectively.

It is well known that General Relativity can be reformulated by using the
tetrad as the dynamical variable. In fact, in the so-called Teleparallel
Equivalent of General Relativity (TEGR) \cite{Hay79,Mal94,Arc04} the action
is built from the object of anholonomity $d\mathbf{e}^{a}$, which reads $d%
\mathbf{e}^{a}=e_{\mu }^{a}~T_{\ \ \rho \nu }^{\mu }~dx^{\rho }\wedge
dx^{\nu }$ in a coordinate basis, where
\begin{equation}
T_{\ \ \rho \nu }^{\mu }\ =\ e_{b}^{\ \mu }(\ \partial _{\rho }e_{\ \nu
}^{b}\ -\ \partial _{\nu }e_{\ \rho }^{b})\ .  \label{eq:torsion}
\end{equation}%
$d\mathbf{e}^{a}$ coincides with the torsion $\mathbf{T}^{a}=d\mathbf{e}^{a}+%
\mathbf{\omega }_{b}^{a}\wedge \mathbf{e}^{b}$ when the spin connection $%
\mathbf{\omega }_{b}^{a}$ is chosen to be zero. Such choice is the
Weitzenb\"{o}ck connection \cite{Wei23}, which also implies that the curvature $%
\mathbf{R}_{b}^{a}=d\mathbf{\omega }_{b}^{a}+\mathbf{\omega }_{c}^{a}\wedge
\mathbf{\omega }_{b}^{c}$ vanishes. According to Eq.~\eqref{eq:torsion} the
Christoffel symbols for the Weitzenb\"{o}ck connection are $\overset{\text{%
{\tiny {W}}}}{\Gamma }\,_{\ \rho \nu }^{\mu }=e_{a}^{\ \mu }\
\partial _{\nu }e_{\ \rho }^{a}=-e_{\ \rho }^{a}\ \partial _{\nu
}e_{a}^{\ \mu }$. Weitzenb\"{o}ck connection leads to the vanishing of
the covariant derivative of the tetrad, which also means that this
connection is metric-compatible (see Eq.~\eqref{eq:orthonormal}).
This property defines an absolute parallelism in the spacetime. In
fact, $\overset{\text{{\tiny {W}}}}{\nabla }e_{a}^{\ \mu }\equiv 0$
means that a parallel transported vector keeps constant its
projections on the tetrad, irrespective of the path as a
consequence of the zero curvature. This is the reason why the name \emph{%
teleparallelism} comes up. As it can be seen, Weitzenb\"{o}ck torsion %
\eqref{eq:torsion} is invariant only under global Lorentz transformations of
the tetrad. So, Weitzenb\"{o}ck geometry selects a preferred global frame
modulo global Lorentz transformations, in spite that the metric is invariant
under local Lorentz transformations.

\subsection{The Teleparallel Equivalent of General Relativity}

The TEGR action is
\begin{equation}
S_{T} \ = \ \frac{1}{2\kappa }\ \int d^{4}x\,\ e\ T \, + \, \int d^{4}x\
\,e\ \mathcal{L}_{matter}\ \ ,  \label{eq:TEGR}
\end{equation}%
where $\kappa =8\pi G$, $e= \det[e_{\ \mu }^{a}]= \sqrt{-g}\ $ and the
\textit{torsion scalar} or \textit{Weitzenb\"{o}ck invariant} $T$ is defined
by the contraction of the torsion tensor (\eqref{eq:torsion} and the \textit{%
superpotential}, that is
\begin{equation}
T\equiv S_{\rho }^{\ \ \mu \nu }\,T_{\ \ \mu \nu }^{\rho }\ ,
\label{eq:scalar}
\end{equation}
where $S_{\rho }^{\ \ \mu \nu }\equiv \frac{1}{2} \left(K_{\ \ \,\rho }^{\mu
\nu }\, +\,T_{\lambda }^{\ \,\lambda \mu }\ \delta _{\rho }^{\nu }\,
-\,T_{\lambda }^{\ \,\lambda \nu }\delta _{\rho }^{\mu } \right) $, with $%
K_{\ \ \,\rho }^{\mu \nu }\equiv\frac{1}{2}\,(T_{\rho }^{\ \,\,\mu \nu
}-T_{\ \ \,\,\rho }^{\mu \nu }+T_{\ \ \,\,\rho }^{\nu \mu }) $ tagged as the
contorsion tensor. Then, the torsion scalar can be expressed by the
following quadratic combination of the components of the torsion tensor
\begin{equation}
T = \frac{1}{4}T_{\rho }^{\ \ \mu \nu }\,T_{\ \ \mu \nu}^{\rho }+\frac{1}{2}%
T_{\rho }^{\ \ \mu \nu }\,T_{\ \ \nu \mu}^{\rho }-T_{\rho }^{\ \ \mu \rho
}\,T_{\ \ \sigma \mu}^{\sigma } \ .  \label{eq:torsiontensor}
\end{equation}

One can compute the Levi-Civita scalar curvature $R$ in terms of the tetrad
by using Eq.~\eqref{eq:orthonormal}; then it is obtained a relation between $%
T$ and $R$ which shows that both the Einstein-Hilbert Lagrangian only
differs from the TEGR Lagrangian in a four-divergence,
\begin{equation}
T\,=\, -R[\mathbf{e}_{a}]\,+\, 2\ e^{-1}\,\partial _{\rho }(e\,T_{\ \ \mu
}^{\mu \ \ \rho })\ .  \label{eq:equivalence}
\end{equation}%
Hence the equations of motion are fully equivalent, showing the equivalence
between GR and TEGR pictures. Although the tetrad field has 16 independent
components, in contrast with the 10 independent components of the symmetric
metric tensor, both TEGR and GR have the same number of degrees of freedom,
as it is expected from their equivalence. In fact, Eq.~\eqref{eq:equivalence}
implies that TEGR only governs the dynamics of the metric, which is
invariant under local Lorentz transformations of the tetrad. Then, the
tetrad is only determined modulo this local symmetry group. Therefore, the
extra components of the tetrad does not represent new dynamical degrees of
freedom. In fact, the local Lorentz invariance of the relation %
\eqref{eq:orthonormal} means that the metric is related with an infinite set
of tetrad fields, connected by local Lorentz transformations. Note also that
teleparallel vacuum solutions do not compel $T$ to vanish, as opposite to $R$
in GR. This point is clearly shown in Eq.~\eqref{eq:equivalence} where $R=0$
implies that $T$ is a four-divergence.

\subsection{$f(T)$ gravity}

\label{sec:lack}

In analogy with $f(R)$ gravity \cite{Buc70, Fel10, Sot10, Cap11, Noj11,
Olm11}, where the GR Lagrangian is extended to an arbitrary function $f$ of
the curvature scalar $R$, $f(T)$ gravity is obtained by replacing the TEGR
Lagrangian with an arbitrary function $f$ of the torsion scalar $T$ \cite%
{Fer07,Fer08,Fer12}
\begin{equation}
S_{f(T)}\ =\frac{1}{2\kappa }\ \int d^{4}x\ e\ f(T)\ \,+\,\int d^{4}x\ e\
\mathcal{L}_{matter}\ .
\end{equation}%
The dynamical equations of $f(T)$ gravity are computed by varying the
modified teleparallel action with respect to the tetrad, yielding
\begin{equation}
4\ e_{a}^{\ \lambda }\ S_{\lambda }^{\ \ \mu \nu }\ \partial _{\mu }T\
f^{\prime \prime }(T)+4\ \left[ e_{a}^{\ \lambda }\ T_{\ \ \mu \lambda
}^{\rho }\ S_{\rho }^{\ \ \mu \nu }+e^{-1}\partial _{\mu }(e\ e_{a}^{\
\lambda }S_{\lambda }^{\ \ \mu \nu })\right] f^{\prime }(T)-e_{a}^{\ \nu }\
f(T)=-2\,\kappa \ e_{a}^{\ \lambda }\ \mathcal{T}_{\lambda }^{\ \nu }\ ,
\label{eq:eom}
\end{equation}%
where $\mathcal{T}_{\lambda }^{\ \nu }$ is the energy-momentum tensor. Of
course, TEGR dynamics is recovered if $f(T)=T$. Remarkably, it is
straightforward to verify that the equations of motion are of second order
with respect to the tetrad (just as in GR with respect to the metric), since
$T_{\ \ \mu \lambda }^{\rho }$ and $S_{\rho }^{\ \ \mu \nu }$ are linear in
first derivatives of the tetrad. Instead, other alternative theories of
gravity have dynamical equations of higher order.

From the equivalence relation expressed in Eq.~\eqref{eq:equivalence}, it is
manifest that $f(T)$ gravity is a non-local Lorentz invariant theory; by
extending the TEGR Lagrangian to a function $f(T)$, the four-divergence
(non-invariant) term remains encapsulated inside the function $f$.\footnote{%
The equivalence between GR and TEGR is broken after the generalization
procedure; therefore $f(T)$ gravity is not at all equivalent to $f(R)$
gravity.}

While TEGR is a theory for the metric (as GR is), $f(T)$ gravity is a theory
for the tetrad. This is so because the modified teleparallel dynamical
equations are not invariant under local Lorentz transformations but only
under global Lorentz transformations.\footnote{%
It should be noted too that the local Lorentz symmetry could be broken at
the level of quantum gravity (Planck scale) and many candidate theories of
quantum gravity predict at some point the loss of local Lorentz symmetry. In
fact, there are many attempts to examine in detail the range in which the
Lorentz symmetry is preserved \cite{Ame04, Mat05,Jac06, Lib13, Blu14, Tas14}%
. Other alternative descriptions of gravity in vogue nowadays such that Ho\u{%
r}ava-Lifshitz \cite{Hor09, Sot11-HL} and {\AE }ther-Einstein \cite{Jac01,
Jac08} are not local Lorentz invariant in the time sector.} Thus the
dynamical equations possess information exclusively associated to the tetrad
field. This means that $f(T)$ theories dynamically endow the spacetime with
an absolute parallelism. This also means that if looking for a solution
associated with a given metric, then the symmetries of the metric are not
enough to anticipate the form of the tetrad solving the equations. For
instance, it was shown that Schwarzschild geometry \cite{Fer11b} and
non-flat FLRW spacetimes \cite{Fer11a} require non-trivial tetrads in order
to consistently solve the equations of motion. Therefore the gravitational
field is encoded in a set of preferred reference frames, which should not
depend on the function $f$ considered \cite{Tam12}. In this line, it was
proposed that a set of suitable frames would be defined up to \emph{certain}
local Lorentz transformations, which are a remnant symmetry group which
depends on the specific spacetime considered \cite{Fer15}. In fact, in a
mathematical context, it has been known for a long time that the vector
fields capable to parallelize a manifold are not unique (see for instance,
Ref.~\cite{topo51}).

In the present work, we rely on results displayed in \cite{Fer11b,Bej15}
where Schwarzschild and Kerr spacetimes proved to remain as vacuum solutions
in $f(T)$ gravity, since both geometries admit a tetrad where the torsion
scalar vanishes ($T=0$) or is constant ($T=T_{c}$). In fact if $\partial
_{\mu }T=0$, Eq.~\eqref{eq:eom} can be arranged as follows
\begin{equation}
G_{\eta }^{\ \nu }\,-\,\frac{1}{2}\,\delta _{\eta }^{\ \nu }\,T_{c}\,+\,%
\frac{1}{2}\,\delta _{\eta }^{\ \nu }\,\frac{f(T_{c})}{f^{\prime }(T_{c})}\
=\ \frac{\kappa }{f^{\prime }(T_{c})}\,\mathcal{T}_{\eta }^{\ \nu }\ ,
\label{eq:Tconst}
\end{equation}%
where we define $G_{\eta }^{\ \nu }$ from Eq.~\eqref{eq:eom} by taking $%
f(T)=T$,
\begin{equation}
G_{\eta }^{\ \nu }\equiv -2e_{\ \eta }^{a}\ \left[ e_{a}^{\ \lambda }\ T_{\
\ \mu \lambda }^{\rho }\ S_{\rho }^{\ \ \mu \nu }\,+\,e^{-1}\,\partial _{\mu
}(e\ e_{a}^{\ \lambda }\ S_{\lambda }^{\ \ \mu \nu })\right] +\frac{1}{2}\
\delta _{\eta }^{\ \nu }\ T\
\end{equation}%
($G_{(\mu \nu )}$ is the Einstein tensor). Remarkably the dynamical
equations \eqref{eq:Tconst}, when the energy-momentum tensor is symmetric,
are nothing but (TEGR) Einstein equations with a scaled Newton constant $%
\tilde{G}=G/f^{\prime }(T_{c})$ and cosmological constant $\Lambda
=(T_{c}-f(T_{c})/f^{\prime }(T_{c}))/2$. Therefore, if $f(T=0)=0$ then the
TEGR solutions having $T=0$ remain as solutions of $f(T)$ gravity (in the
case of non-vacuum solutions we should adjust the Newton constant). Even if $%
T_{c}\neq 0$ one could still regard any TEGR solution having $T=T_{c}$ as a
solution to $f(T)$ equations for proper values of cosmological and Newton
constants. In the following sections we will take advantage of this property
to figure out whether a metric solving Einstein's equations could survive as
a solution of $f(T)$ equations or not. Following the strategy displayed in
\cite{Bej15}, we will exploit the local Lorentz invariance of TEGR to look
for solutions having $T=0$ by starting from a known solution. If we success,
then we will state that the so obtained solution remains as a solution to $%
f(T)$ gravity. The null tetrad approach will be very useful for this purpose.


\section{McVittie geometry}

\label{sec:mcv-geo}

McVittie geometry \cite{McVittie} describes a cosmological black hole, which
means a black hole solution embedded in an expanding FLRW universe. A
comprehensive review of McVittie geometry was carried out by Nolan in the
late nineties \cite{Nolan}. Nonetheless, Kaloper et al. \cite{Kaloper} went
one step further by analyzing some misconceptions in the geometrical
interpretation of the solution. The original McVittie spacetime assumes
vanishing (Levi-Civita) spatial curvature in the asymptotically FLRW region,
but also it can be easily generalized to include positive or negative
spatial curvature \cite{Sakai}. Here we will focus on the spatially flat
case. It is assumed that the spatial curvature of the FLRW does not
significantly influence the dynamics around the central mass, accounting for
that the radius of curvature is (much) greater than the gravitational radius
of the mass source. Then, the standard McVittie geometry is described by the
metric
\begin{equation}
ds^{2}=\left( \frac{1-\mu }{1+\mu }\right) ^{2}dt^{2}-(1+\mu )^{4}a^{2}(t)~d%
\mathbf{x}^{2},  \label{eq:old-mcv}
\end{equation}%
where $\mu =m(2a(t)|\mathbf{x}|)^{-1}$, with $m/G$ the mass of the source, $%
a(t)$ is the asymptotic cosmological scale factor, and the center of the
spherical symmetry is at $\mathbf{x}=0$. This is an exact solution to
Einstein's equations for an arbitrary mass $m$ provided that $a(t)$ solves
the Friedmann equation. As might be expected, if $m=0$ the geometry
corresponds to the standard flat FLRW spacetime, and if $a(t)=1$ the line
element describes the Schwarzschild solution (in isotropic coordinates). For
$a\sim e^{H_{0}t}$, the metric reduces to the case of Schwarzschild-de
Sitter (by adopting a positive cosmological constant). It can be verified
that McVittie solution has a spacelike and inhomogeneous singularity at $\mu
=1$ (which means that $a(t)|\mathbf{x}|=m/2$) where the surface lies in the
causal past of all spacetime events so that it should be properly
interpreted as a cosmological big-bang singularity (see Ref.~\cite{Kaloper}
for more detail).

In analogy to the FLRW case for a perfect fluid, the energy density scales
with the scale factor $a(t)$ and commands the expansion rate
\begin{equation}
\rho =\frac{3H^{2}(t)}{\kappa }\ ,
\end{equation}%
where $H(t)=\dot{a}(t)/a(t)$ is the Hubble parameter. Remarkably, the energy
density is constant along slices where $t$ is constant, but the pressure on
fixed $t$ slices is not homogeneous
\begin{equation}
p=\frac{1}{\kappa }\left( -3~H(t)^{2}+2~\dot{H}^{2}(t)\left( \frac{m+2~|%
\mathbf{x}|~a(t)}{m-2~|\mathbf{x}|~a(t)}\right) \right) \ .
\end{equation}%
The inhomogeneous pressure constitutes the necessary non-gravitational
balancing force (when the mass is constant and the energy density is
spatially homogeneous) to compensate the gravitational attraction of the
central mass.

The coordinates employed in the line element given by Eq.~\eqref{eq:old-mcv}
are such that $|\mathbf{x}|$ covers the exterior of the black hole twice,
that is $m/2<|\mathbf{x}|<\infty $ covers the same exterior region as $0<|%
\mathbf{x}|<m/2$. Therefore, Kaloper et al. \cite{Kaloper} propose another
coordinate choice which in turn imitates better the familiar static form of
the Schwarzschild metric. The new radial coordinate is defined by
\begin{equation}
\mathbf{R}=(1+\mu )^{2}\,a(t)\,\mathbf{x}\ ,  \label{eq:rmuax}
\end{equation}%
where $|\mathbf{R}|=R$ represents the \textquotedblleft spherical
area\textquotedblright\ coordinate. Since the relation between $a(t)|\mathbf{%
x}|$ and $R$ is quadratic, the coordinate transformation \eqref{eq:rmuax}
actually defines two separate branches
\begin{equation}
a(t)~|\mathbf{x}|=\frac{m}{2}\left( \frac{R}{m}-1\pm \sqrt{\left( \frac{R}{m}%
-1\right) ^{2}-1}\right) ^{-1}\ .
\end{equation}%
The physically relevant branch is the one with the ($-$) sign, since $%
R\rightarrow \infty $ implies $|\mathbf{x}|\rightarrow \infty $, therefore
the geometry is asymptotically FLRW-like.

By applying the transformation defined in \eqref{eq:rmuax}, the McVittie
metric becomes
\begin{equation}
ds^{2}=\left( 1-\dfrac{2m}{R}-H(t)^{2}R^{2}\right) dt^{2}+\frac{2H(t)R}{%
\sqrt{1-2m/R}}\ dR\ dt-\frac{dR^{2}}{1-2m/R}-R^{2}d\Omega ^{2}\ ,
\label{eq:nsmcv}
\end{equation}%
where $d\Omega^2 =d\theta^2+\sin^2\theta\,d\phi^2$. It is now evident that a
constant value of $H$ leads to the Schwarzschild-de Sitter metric in
coordinates which are analogous to outgoing Eddington-Finkelstein
coordinates.


\section{McVittie solution in $f(T)$ gravity}

\label{sec:mcv-sol}

As previously mentioned, $f(T)$ gravity is not locally Lorentz invariant.
This means that tetrads connected by local Lorentz transformations, which
reproduce the same metric tensor, are not equivalent at the level of the
equations of motion of $f(T)$: not all of them will represent a proper
parallelizing field of frames (in the sense of being a consistent solution
of the dynamical equations). We can state that $f(T)$ gravity selects the
parallelization of the spacetime. Finding suitable tetrad solutions is then
quite awkward in $f(T)$ theory, since the symmetry of the searched metric is
not enough to determine the form of the tetrad. However, as explained at the
end of Section \ref{sec:tele}, one can exploit the local Lorentz invariance
of TEGR solutions to force the scalar torsion $T$ to be zero. 
If such a purpose is attainable, we will obtain a tetrad solving $f(T)$ gravity
equations too (whenever $f(T=0)\neq 0$). This also means that the same
metric solving TEGR equations is admisible in $f(T)$ gravity. Since the
condition $T=0$ is not affected by global linear transformations of the
tetrad, it could be easier to look for the condition $T=0$ by using a null
tetrad \cite{Bej15}: given an orthonormal tetrad $\{\mathbf{e}^{a}\}=\{%
\mathbf{e}^{0},\mathbf{e}^{1},\mathbf{e}^{2},\mathbf{e}^{3}\}$, a null
tetrad can be defined as
\begin{equation}
\{\mathbf{n}^{a}\}=\{\mathbf{l},\mathbf{n},\mathbf{m},\overline{\mathbf{m}}%
\}=\left\{ \frac{(\mathbf{e}^{0}-\mathbf{e}^{1})}{\sqrt{2}},\frac{(\mathbf{e}%
^{0}+\mathbf{e}^{1})}{\sqrt{2}},\frac{(\mathbf{e}^{2}+i\,\mathbf{e}^{3})}{%
\sqrt{2}},\frac{(\mathbf{e}^{2}-i\,\mathbf{e}^{3})}{\sqrt{2}}\right\} \ .
\label{eq:null}
\end{equation}%
$\{\mathbf{n}^{a}\}$ is a null basis, $\mathbf{n}^{a}\cdot \mathbf{n}^{a}=0$%
, but it is not orthogonal ($\mathbf{l}\cdot \mathbf{n}=-\mathbf{m}\cdot
\overline{\mathbf{m}}=1$). Then, Eq.~\eqref{eq:orthonormal} can be rewritten
in this new basis as
\begin{equation}
g_{\mu \nu }\ =\ \eta _{ab}\,n_{\mu }^{a}\,n_{\nu }^{b}~,
\end{equation}%
where $\eta _{ab}$ now reads
\begin{equation}
\eta _{ab}=\left(
\begin{array}{cccc}
\ 0 & \ 1 & \ 0 & \ 0 \\
\ 1 & \ 0 & \ 0 & \ 0 \\
\ 0 & \ 0 & \ 0 & -1 \\
\ 0 & \ 0 & -1 & \ 0%
\end{array}%
\right) \ .
\end{equation}%
In terms of the tensorial product of the elements of the null tetrad the
metric reads $\mathbf{g}=\mathbf{n}\otimes \mathbf{l}+\mathbf{l}\otimes
\mathbf{n}-\mathbf{m}\otimes \overline{\mathbf{m}}-\overline{\mathbf{m}}%
\otimes \mathbf{m}$. A local Lorentz boost along the direction of $\mathbf{e}%
^{1}$, with parameter $\gamma (x^{\mu })=\cosh [\lambda (x^{\mu })]$, is the
transformation $\{\mathbf{l},\mathbf{n}\}\mapsto \{\exp [-\lambda (x^{\mu
})]\,\mathbf{l},\exp [\lambda (x^{\mu })]\,\mathbf{n}\}$, which clearly does
not modify the form of the metric tensor. The null tetrad approach to get $%
T=0$ exploits the freedom in the choice of the function $\lambda (x^{\mu })$%
. The strategy is the following: (i) determine the null tetrad $\{\mathbf{n}%
^{a}\}$, (ii) apply the transformation in the $\{\mathbf{l},\mathbf{n}\}$%
-sector, (iii) impose the condition that the torsion scalar $T$ be zero.
This approach was successfully implemented for the Kerr geometry \cite{Bej15}%
; here we will show that it is also useful in McVittie spacetime.\footnote{%
In both cases, the success of the null tetrad approach seems to reside in
the fact that the radial boost does not mix the sector $\{t,r\}$ with the
angular part $\{\theta ,\phi \}$.}

Then, we compute the null tetrad associated to the metric given in Eq.~%
\eqref{eq:nsmcv} with coordinates $(t,R,\theta ,\phi )$ and we perform a
radial boost yielding
\begin{equation}
n_{\, \, \mu }^{a}=\dfrac{1}{\sqrt{2}}\left(
\begin{array}{cccc}
e^{-\lambda (t,R,\theta)}\left( \sqrt{1-2m/R}+RH(t)\right) & -e^{-\lambda
(t,R,\theta)} \frac{1}{\sqrt{1-2m/R}} & 0 & 0 \\
e^{\lambda (t,R,\theta)}\left( \sqrt{1-2m/R}-RH(t)\right) &
e^{\lambda(t,R,\theta)}\frac{1}{\sqrt{1-2m/R}} & 0 & 0 \\
0 & 0 & R & iR\sin\theta \\
0 & 0 & R & -iR\sin\theta%
\end{array}%
\right) \ ,  \label{eq:nulltetrad}
\end{equation}%
where the function $\lambda$ has to be determined. For this null tetrad, we
determine the torsion scalar by means of Eq.~\eqref{eq:scalar}:
\begin{equation}
T\,=-6\, H(t)^{2}\,+\,2\, R^{-2}\, -\, 4\, R^{-1}\, \partial _{t}\lambda\ .
\end{equation}%
By imposing that the torsion scalar vanishes, we solve the differential
equation to obtain
\begin{equation}
\lambda(t,R)=\dfrac{t}{2R}-\dfrac{3R}{2}\int H^{2}(t)\ dt.  \label{eq:lambda}
\end{equation}%
Note that $\lambda$ allows for an additive function of $(R, \theta)$ without
affecting the result $T=0$.

In summary, according to Eq.~\eqref{eq:Tconst}, we conclude that $f(T)$
gravity is not able to deform McVittie metric since we have just found a
solution leading to a vanishing torsion scalar which consistently solves the
dynamical equations (notice the scaling of Newton constant $\kappa $),
\begin{eqnarray}
2\kappa \rho  &=&f(T=0)+6~H^{2}(t)~f^{\prime }(T=0)=6~H^{2}(t)~f^{\prime
}(T=0)~,  \label{EOM_McV_rho} \\
2\kappa (p+\rho ) &=&-4~f^{\prime }(T=0)~\dot{H}(t)~\frac{1}{\sqrt{1-2m/R}}\
,  \label{EOM_McV_p}
\end{eqnarray}%
where, due to the non-diagonal form of the metric tensor \eqref{eq:nsmcv},
the stress-energy tensor for a perfect fluid becomes
\begin{equation}
T_{\ \ \nu }^{\mu }=\left(
\begin{array}{cccc}
\rho  & (\rho +p)~R~H(t)\sqrt{1-2m/R} & 0 & 0 \\
0 & -p & 0 & 0 \\
0 & 0 & -p & 0 \\
0 & 0 & 0 & -p%
\end{array}%
\right) ,  \label{eq:tem-mcv}
\end{equation}%
with $\rho $ and $p$ the energy density and the pressure, respectively \cite%
{Sakai}.


\section{Cosmology with $T=0$}

\label{sec:cosmo}

As it has been extensively studied in the literature, flat FLRW cosmology
accepts the naive diagonal tetrad (in the Cartesian chart) as a suitable
solution in the context of $f(T)$ gravity, which leads to $T=-6H^{2}(t)$
\cite{Fer07, Fer11a}. However, McVittie metric for $m=0$ reduces to FLRW
metric. Therefore, we have also obtained the outstanding result that there
exists a a tetrad $\{\mathbf{e}^{a}\}$ having $T=0$ in flat FLRW spacetime,
from which immediately follows that that such tetrad is also a solution $f(T)
$ gravity (assuming $f(T=0)\neq 0$, and taking care of the scaling of Newton
constant). By replacing $m=0$ in Eq.~\eqref{eq:nulltetrad}, we get that the
null tetrad associated with $\{\mathbf{e}^{a}\}$ is
\begin{equation}
n_{\,\,\mu }^{a}=\dfrac{1}{\sqrt{2}}\left(
\begin{array}{cccc}
e^{-\lambda (t,R)}\left( 1+RH(t)\right)  & -e^{-\lambda (t,R)} & 0 & 0 \\
e^{\lambda (t,R)}\left( 1-RH(t)\right)  & e^{\lambda (t,R)} & 0 & 0 \\
0 & 0 & R & iR\sin \theta  \\
0 & 0 & R & -iR\sin \theta
\end{array}%
\right) \ ;  \label{enula-cosmo}
\end{equation}%
where $\lambda (t,R)$ is the function given in Eq.~\eqref{eq:lambda}.
According to Eq.~\eqref{EOM_McV_p}, the dynamical equations for $m=0$ become
\begin{eqnarray}
2\kappa \rho  &=&f(T=0)+6~H^{2}(t)~f^{\prime }(T=0)=6~H^{2}(t)~f^{\prime
}(T=0)\ , \\
2\kappa (p+\rho ) &=&-4~f^{\prime }(T=0)~\dot{H}(t)\ ,
\end{eqnarray}
This remarkable result will be deeply studied in a forthcoming article \cite%
{bgff:2016}. At first sight, solutions providing the same metric tensor but
different torsion scalars suggest the involvement of the extra degrees of
freedom characteristic of $f(T)$ gravity. They constitute differrent
admisible parallelizations of FLRW geometry.

\bigskip

To compare both $f(T)$ solutions, firstly, we will get the orthonormal
tetrad $e_{\ \mu }^{a}$ associated with the null tetrad given in Eq.~%
\eqref{enula-cosmo}, and then we will write it in the original chart $(t,|%
\mathbf{x}|)$. In fact, by performing the inverse coordinate transformation
of Eq.~\eqref{eq:rmuax}, which reduces to $R=a(t)~|\mathbf{x}|=a(t)~r$ for $%
m=0$, we will get $e_{\ \mu ^{\prime }}^{a}$, in terms of the spherical radial
coordinate $r$. The first step is achieved through a transformation $L_{\
b}^{a}$ such that (see Eq.~\eqref{eq:null})
\begin{equation}
\begin{split}
e_{\ \mu }^{a}\ =\ L_{\ b}^{a}~n_{\,\,\mu }^{b}& =\dfrac{1}{\sqrt{2}}\left(
\begin{array}{cccc}
1 & 1 & 0 & 0 \\
-1 & 1 & 0 & 0 \\
0 & 0 & 1 & 1 \\
0 & 0 & -i & i%
\end{array}%
\right) \times \left(
\begin{array}{cccc}
e^{-\lambda }\left( 1+RH(t)\right)  & -e^{-\lambda } & 0 & 0 \\
e^{\lambda }\left( 1-RH(t)\right)  & e^{\lambda } & 0 & 0 \\
0 & 0 & R & i\,R\,\sin \theta  \\
0 & 0 & R & -i\,R\,\sin \theta
\end{array}%
\right)  \\
& =\left(
\begin{array}{cccc}
\cosh \lambda -R\,H(t)\sinh \lambda \  & \sinh \lambda  & 0 & 0 \\
\sinh \lambda -R\,H(t)\cosh \lambda \  & \cosh \lambda  & 0 & 0 \\
0 & 0 & R & 0 \\
0 & 0 & 0 & R\,\sin \theta
\end{array}%
\right) \ ,
\end{split}%
\end{equation}%
where we set $\lambda \equiv \lambda (t,R)$ to abbreviate the notation. In
the second step, we will perform the coordinate transformation $x^{\mu
}=(t,R,\theta ,\phi )$ $\longrightarrow $ $x^{\mu ^{\prime }}=(t,r,\theta
,\phi )$. The tetrad $\{\mathbf{e}^{a}\}$ is a geometrical object
independent of the coordinate choice,
\begin{equation}
\mathbf{e}^{a}=e_{\ \mu }^{a}~dx^{\mu }=e_{\ \mu ^{\prime }}^{a}~dx^{\mu
^{\prime }},
\end{equation}%
but its components $e_{\ \mu }^{a}$ will change. According to Eq.~%
\eqref{eq:rmuax}, the 1-forms $dx^{\mu }$ and $dx^{\mu ^{\prime }}$ will not
change except for $dR=\dot{a}~r~dt+a~dr$. Then, it is obtained that
\begin{equation}
e_{\ \mu ^{\prime }}^{a}=\left(
\begin{array}{cccc}
\cosh \lambda  & a(t)\,\sinh \lambda  & 0 & 0 \\
\sinh \lambda  & a(t)\,\cosh \lambda  & 0 & 0 \\
0 & 0 & a(t)r & 0 \\
0 & 0 & 0 & a(t)\,r\,\sin \theta
\end{array}%
\right) \ .  \label{eq:nodiagcosmo}
\end{equation}%
Tetrad \eqref{eq:nodiagcosmo} is the one having $T=0$, as written in usual
spherical coordinates. On the other hand, there exist a tetrad which has $%
T=-6H^{2}$; this tetrad is diagonal in the Cartesian chart, but in spherical
coordinates reads
\begin{equation}
e_{\ \mu ^{\prime }}^{a^{\prime }}=\left(
\begin{array}{cccc}
1 & 0 & 0 & 0 \\
0 & a(t)\,\sin \theta \,\cos \varphi  & a(t)\,r\,\cos \theta \,\cos \varphi
& -a\,(t)\,r\,\sin \theta \,\sin \varphi  \\
0 & a(t)\,\sin \theta \,\sin \varphi  & a(t)\,r\,\cos \theta \,\sin \varphi
& a(t)\,r\,\sin \theta \,\cos \varphi  \\
0 & a(t)\,\cos \theta  & -a(t)\,r\,\sin \theta  & 0%
\end{array}%
\right) \ .  \label{eq:cosmocartesf}
\end{equation}%
Of course it should exist a local Lorentz transformation connectin the
tetrads  \eqref{eq:nodiagcosmo} and \eqref{eq:cosmocartesf}, since both
describre the same FLRW metric. In fact, the local Lorentz transformation is%
\begin{equation}
\Lambda _{~~b}^{a^{\prime }}=\left(
\begin{array}{cccc}
\cosh \lambda  & \sinh \lambda  & 0 & 0 \\
\sinh \lambda \,\sin \theta \,\cos \phi  & \cosh \lambda \,\sin \theta
\,\cos \phi  & \cos \theta \,\cos \phi  & -\sin \phi  \\
\sinh \lambda \,\sin \theta \,\sin \phi  & \cosh \lambda \,\sin \theta
\,\sin \phi  & \cos \theta \,\sin \phi  & \cos \phi  \\
\sinh \lambda \,\cos \theta  & \cosh \lambda \,\cos \theta  & -\sin \theta
& 0%
\end{array}%
\right) ,  \label{eq:lorentz}
\end{equation}%
i.e. $\mathbf{e}^{a^{\prime }}=\Lambda _{~~b}^{a^{\prime }}~\mathbf{e}^{b}$,
which separates as the product of two rotations and one boost
\begin{equation}
\Lambda _{~~b}^{a^{\prime }}=\left(
\begin{array}{cccc}
1 & 0 & 0 & 0 \\
0 & 0 & \cos \phi  & -\sin \phi  \\
0 & 0 & \sin \phi  & \cos \phi  \\
0 & 1 & 0 & 0%
\end{array}%
\right) \left(
\begin{array}{cccc}
1 & 0 & 0 & 0 \\
0 & \cos \theta  & -\sin \theta  & 0 \\
0 & \sin \theta  & \cos \theta  & 0 \\
0 & 0 & 0 & 1%
\end{array}%
\right) \left(
\begin{array}{cccc}
\cosh \lambda  & \sinh \lambda  & 0 & 0 \\
\sinh \lambda  & \cosh \lambda  & 0 & 0 \\
0 & 0 & 1 & 0 \\
0 & 0 & 0 & 1%
\end{array}%
\right) .
\end{equation}


In Ref.~\cite{Fer15} it was shown that each solution to $f(T)$ dynamical
equations can allow for a set of local Lorentz remnant symmetries. Such
symmetries are generated by Lorentz transformations accomplishing the
condition
\begin{equation}
d(\epsilon _{abcd}\ \mathbf{e}^{a}\wedge \mathbf{e}^{b}\wedge \eta
^{de}\Lambda _{\ f^{\prime }}^{c}d\Lambda _{\ e}^{f^{\prime }})=0~.
\end{equation}%
Our pair $(\mathbf{e}^{a},\Lambda _{\ b}^{a^{\prime }})$ does not satisfy
this relationship. This is because, although both tetrads have consistent
equations of motion, their associated torsion scalars are different; so the
transformation \eqref{eq:lorentz} is not a symmetry of the action. Also,
they have different classification concerning to the \textit{%
n-closed-area-frame} (n-CAF), distinction firstly introduced in the same
paper. Remember that a solution of $f(T)$ gravity is $n$-CAF if $n$ of the
six pairs $(\mathbf{e}^{a},\mathbf{e}^{b})$ satisfy the equation
\begin{equation}
d(\mathbf{e}^{a}\wedge \mathbf{e}^{b})=0.
\end{equation}%
In this particular case, it can be proven that the tetrad %
\eqref{eq:nodiagcosmo} is a 1-CAF, since the only combination that is zero
is the one with $(\mathbf{e}^{0},\mathbf{e}^{1})$. On the other hand, the
tetrad \eqref{eq:cosmocartesf} is a 3-CAF \cite{Fer15}. This means that the
second one allows for three independent \textit{local} Lorentz
transformations leaving $T$ unchanged, while the first solution admits only
the local boost associated with the remnant freedom of the function $\lambda$%
. On the contrary, the transformation \eqref{eq:lorentz} is not a remnant
symmetry of the $f(T)$ dynamical equations because it changes the torsion
scalar $T$.


\section{Conclusions}

The main purpose of $f(T)$ gravity would be finding low/high-energy
deformations of Einstein gravity to solve its shortcomings in a geometric
framework. However, the null tetrad approach shows that it is rather easy to
get TEGR solutions with constant torsion scalar. These solutions survive in $%
f(T)$ gravity, associated with modified Newton and cosmological constants.
This is the way we followed to show that McVittie geometry is a solution to $%
f(T)$ gravity. Remarkably, by taking $m=0$ we also obtained a new
consistent solution for FLRW universe in $f(T)$ gravity, which has
$T=0$. To the best of our knowledge, this is the first time that a
consistent cosmological solution with a vanishing torsion scalar is
introduced in the literature. The fact that the torsion scalar
differs from $-6H^{2}$ could be a manifestation of the extra degrees
of freedom of the theory, which is right now a topic under
consideration \cite{bgff:2016}.

\label{sec:disc}


\begin{acknowledgments}
This work was supported by Consejo Nacional de Investigaciones
Cient\'{\i}ficas y T\'{e}cnicas (CONICET) and Universidad de Buenos Aires.
The authors thank to F. Fiorini and N. Vattuone for helpful
discussions.
\end{acknowledgments}


\end{document}